\documentclass[12pt,a4paper]{article}
\usepackage{graphicx}
\begin{document}
\textwidth=135mm
 \textheight=200mm
\begin{center}
{\bfseries Geoneutrinos and Hydridic Earth (or primordially Hydrogen-Rich Planet)
\footnote{{\small Talk at the The International Workshop on Prospects of Particle Physics: «Neutrino Physics and Astrophysics», 
 Resort Hotel «Valday», Valday,  Novgorod region, Russia, 26 January-2 February 2014.}}}
\vskip 5mm
L. Bezrukov$^\dag$ and V. Sinev$^\dag$
\vskip 5mm
{\small {\it $^\dag$ Institute for
Nuclear Research of Russian Academy of Sciences, Moscow, Russia}}
\\
\end{center}
\vskip 5mm

\centerline{\bf Abstract}
Geoneutrino is a new channel of information about geochemical composition of the Earth.  We alnalysed here the following problem. What statistics do we need to distinguish between predictions of Bulk Silicate Earth model and Hydridic Earth model  for Th/U signal ratio? We obtained the simple formula for estimation of error of Th/U signal ratio. Our calculations show that we need more than $22 kt \cdot year$ exposition for Gran-Sasso underground laboratory and Sudbury Neutrino Observatory. We need more than $27 kt \cdot year$ exposition for Kamioka site in the case of stopping of all Japanese nuclear power plants. 
\vskip 10mm

\section{\label{sec:intro}Introduction}
\hspace{0.5cm}Geoneutrino is a new channel of information about geochemical composition of the Earth. Geoneutrino is antineutrino emitted in a decay chain of U,Th and $^{40}$K located in the Earth's interior. The first direct measurement of geoneutrino flux was made by the Borexino collaboration \cite{Borex2013} and the KamLand collaboration \cite{KamLand2013}.  Number of events in these detectors depends on the uranium mass in the Earth $m(U)$, the thorium mass in the Earth $m(Th)$ and on their distribution in the Earth.  The Bulk Silicate Earth (BSE) model \cite{Mantovani2003} gives  $m_{BSE}(U) = 0,81 \cdot 10^{17}kg, m_{BSE}(Th) = 3,16 \cdot 10^{17}kg, m_{BSE}(K) = 0.49 \cdot 10^{21}kg $. This amount distributes only in Crust and Upper Mantel in the frame of BSE model.
Basic idea of BSE model  \cite{Mantovani2003} is that the
Earth chemical composition must be the same as meteorite chemical composition.  The meteorites come mostly from Asteroid Belt (AB).  So, the Earth chemical composition must be the same as AB chemical composition. 

Chondritic ratio is one of the main characteristic of AB chemical composition and varies from 2.6 to 4.2  \cite{lover}, \cite{shinot}.  The average value for solar system is proposed to be:

\begin{equation}\label{eq:1}
R(\mbox{Th/U})_{AB} = m_{Th}/m_U  = 3,9
\end{equation}

The asteroid belt is the region of the Solar System located roughly between the orbits of the planets Mars and Jupiter. Some of the debris from collisions can form meteoroids that enter in the Earth's atmosphere. Of the 50,000 meteorites found on Earth to date, 99.8 percent are believed to have originated in the asteroid belt.

There is the alternative Earth model \cite{Larin}, \cite{Larin2012} named Hydridic Earth model (HE) which predicts the primordial chemical elements composition of the Earth. The basic idea of this model is the dependence of planet chemical composition on the distance from the Sun.

Vladimir Larin \cite{Larin} used the idea that the separation of the chemical elements in the solar system (chemical differentiation) was originated from the magnetic
field of the Protosun. He found a correlation between the ratio of the Earth crust chemical element abundances to Sun chemical
element abundances and the first ionization potential of these elements.
The observed correlation is theoretically \cite{Larin2012} interpreted as a Boltzmann distribution. The numerical model was succesfully tested for the observed solar normalized chemical compositions of the Earth, Mars and chondrites. 

The 18.3\% of the Earth primordial mass is predicted to be Hydrogen  \cite{Larin2012}. 
The inner Earth would have been and still
could be hydrogen rich. The most part of primordial hydrogen have
escaped to atmosphere and space through the degassing of the mantle. Model suggests that large amounts of hydrogen are still 
located in the core.

\hspace{0.5cm}On the base of HE model the work  \cite{bezruk} calculated U, Th and $^{40}$K abundances in the Earth: 
 $m(U) = 3.15 \cdot 10^{17} kg$,
$m(Th) = 5.42 \cdot 10^{17} kg$, 
$m(^{40}K) = 2.63 \cdot 10^{19} kg$
and obtained Th/U mass ratio for the Earth: 

\begin{equation}\label{eq:2}
R(\mbox{Th/U})_{HE} =\left(\frac{m_{Th}}{m_{U}}\right)_{HE} = 1.72.
\end{equation}
 
 This value is different from chondritic Th/U mass ratio of 3.9 usually used.  The accurate measurement of this ratio could permit to choose between BSE model and HE model. 

The ability of discrimination between HE ans BSE models is limited not only by experimental uncertainty but also by uncertainty of theoretical predictions. The main uncertainty arises from the unknown distribution of Th and U concentrations in the Earth interior. The prediction of Th/U signal ratio is free from this uncertainty. 
 
We alnalysed here the following problem. What statistics and what level of background must the geoneutrino detector have for discrimination between predictions of Bulk Silicate Earth model and Hydridic Earth model?
\vskip 10mm

\section{Counting rate of events in geoneutrino detector}

The detector can record the geoneutrino from U and Th decays through the reaction of inverse beta decay:
\begin{equation}\label{eq:32}
\tilde{\nu} + p = e^+ + n
\end{equation}
This reaction has threshould is equal to 1.806 MeV. 

 We calculated  for BSE model the counting rate of events in geoneutrino detector from thorium and uranium decays separately and from nuclear reactors as background. The results are shown for $1 kt \cdot year$ exposition in Table 1 and Fig. 1 for detector consisting from C$_n$H$_{2n}$ scintillator and locating at Gran Sasso site.  In Table 2 the results are shown for same detector but locating at Sudbury site and in Table 3 at KamLAND site. We used the programs written by V. Sinev and results described in \cite{Sinev}. We calculated the number of events of reactor atineutrinos for Kamioka site in the case of stopping of all Japanese nuclear power plants and in the case of running. 

\vspace{0.6cm}
\begin{tabular}{ccc}
\multicolumn{3}{c}{{\bf Table 1.} Number of events for $1 kt \cdot year$ exposition
for Gran Sasso}\\
&E = 1,5 - 2,5 MeV& E = 1,0 - 1,5 MeV \\
 &$S_{T,2} = 40$&$S_{T,1} = 29$\\
 &$S_{R,2} = 21$ &$S_{R,1} = 3$\\
 &$S_{U,2} = 19$ &$S_{U,1} = 16$\\
 & &$S_{Th} = 10$ \\
\end{tabular}

\vspace{0.6cm}
\begin{tabular}{ccc}
\multicolumn{3}{c}{{\bf Table 2.} Number of events for $1 kt \cdot year$ exposition
for Sudbury}\\
&E = 1,5 - 2,5 MeV& E = 1,0 - 1,5 MeV \\
 &$S_{T,2} = 61$&$S_{T,1} = 38$\\
 &$S_{R,2} = 37$ &$S_{R,1} = 6$\\
 &$S_{U,2} = 24$ &$S_{U,1} = 20$\\
 & &$S_{Th} = 12$ \\
\end{tabular}

\vspace{0.6cm}
\begin{tabular}{ccc}
\multicolumn{3}{p{13cm}}{{\bf Table 3.} Number of events for $1 kt \cdot year$ exposition for KamLAND in the case of stopping of all Japanese nuclear power plants.}\\
&E = 1,5 - 2,5 MeV& E = 1,0 - 1,5 MeV \\
 &$S_{T,2} = 27$&$S_{T,1} = 23$\\
 &$S_{R,2} = 12$ &$S_{R,1} = 2$\\
 &$S_{U,2} = 15$ &$S_{U,1} = 13$\\
 & &$S_{Th} = 8$ \\
\end{tabular}

\vspace{0.6cm}
\begin{tabular}{ccc}
\multicolumn{3}{p{13cm}}{{\bf Table 4.} Number of events for $1 kt \cdot year$ exposition for KamLAND in the case of running of all Japanese nuclear power plants.}\\
&E = 1,5 - 2,5 MeV& E = 1,0 - 1,5 MeV \\
 &$S_{T,2} = 189$&$S_{T,1} = 47$\\
 &$S_{R,2} = 174$ &$S_{R,1} = 26$\\
 &$S_{U,2} = 15$ &$S_{U,1} = 13$\\
 & &$S_{Th} = 8$ \\
\end{tabular}

where:

E - energy release in the first flash after neutrino reaction in detector,

 $S_{T,2}$ - total number of events in energy range $E = 1,5 - 2,5 MeV$,

$S_{R,2}$ - number of events from reactors in energy range $E = 1,5 - 2,5 MeV$,

$S_{U,2}$ - number of geoneutrino events from U decay in energy range $E = 1,5 - 2,5 MeV$,

$S_{T,1}$ - total number of events in energy range $E = 1,0 - 1,5 MeV$,

$S_{R,1}$ - number of events from reactors in energy range $E = 1,0 - 1,5 MeV$,

$S_{U,1}$ - number of geoneutrino events from U decay in energy range $E = 1,0 - 1,5 MeV$,

$S_{Th}$ - number of geoneutrino events from Th decay in energy range $E = 1,0 - 1,5 MeV$.
\vskip 2mm

\begin{figure}[ht] 
\vspace{10pt}
\label{fig:f1} 
\centering
\includegraphics[scale=1.2]{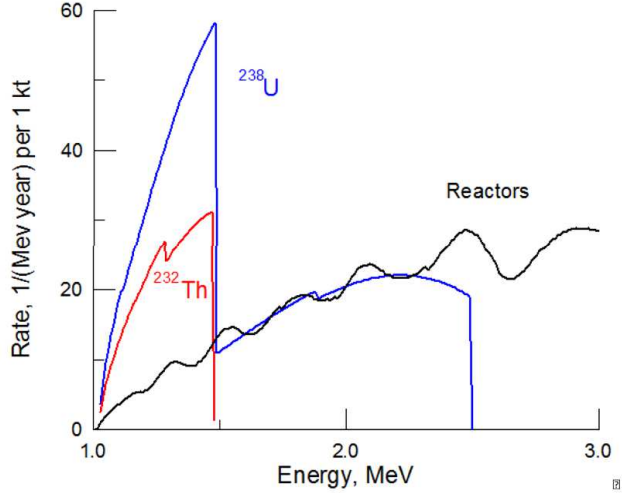}
\caption{Calculated dependence of counting rate of geoneutrino inverse beta decay reactions in 1 kt scintillation detector per year versus energy release in the first flash after neutrino reaction in detector. Blue curve - geoneutrino from U decay, red curve - geoneutrino from Th decay, black curve - calculated background from reactors for Gran-Sasso location.}
\end{figure}

We have from figures of Table 1 for the signal ratio $\left(\frac{S_{Th}}{S_U}\right)_{BSE}$:

\begin{equation}\label{eq:4}
\left(\frac{S_{Th}}{S_U}\right)_{BSE} = 0.28.
\end{equation}

This signal ratio is propotional to R.
So we can calculate this ratio for value (\ref{eq:2}) $R_{HE} = 1.72$: 

\begin{equation}\label{eq:5}
\left(\frac{S_{Th}}{S_U}\right)_{HE} = 0.28 \cdot \frac{1.72}{3.9} = 0.12.
\end{equation}
\vskip 10mm

To discriminate the difference between (\ref{eq:4}) and  (\ref{eq:5}) the experimental accuracy should be better than: 

\begin{equation}\label{eq:41}
\delta\frac{S_{Th}}{S_U}(m, t, \eta)
<\frac{1}{3}\cdot\left(\left(\frac{S_{Th}}{S_U}\right)_{BSE} - \left(\frac{S_{Th}}{S_U}\right)_{HE}\right),
\end{equation}

where $\delta\frac{S_{Th}}{S_U}(m, t, \eta)$ - the error of obtained Th/U signal ratio by geoneutrino detector with fiducial mass $m$ and with efficiency $\eta$ during the operational time $t$.

\section{Evaluation of  Th/U signal ratio}

 Detector measuring antineutrino spectra from geoneutrinos can see the total spectrum generating by uranium and thorium isotopes. But because the fact that uranium spectrum extends to higher energies (Fig.1) we have possibility to separate spectra of uranium and thorium. To do this we need to measure accurately high energy U spectrum part then to restore the low energy part of   U  spectrum and after to subtract it from the total spectrum.

The procedure of  evaluation of values $S_{Th}$, $S_U$  from the experimental data is the following.

$S_{U,2} = S_{T,2} - S_{R,2}$ where $S_{R,2}$ is calculated for the place of detector location. The current accuracy of such calculations is 3\%.

$S_{U,2}$ extrapolats to energy region $E = 1,0 - 1,5 MeV$ by formula

$S_{U,1} = \alpha \cdot S_{U,2}$ where $\alpha$ is calculated and depends on U distribution in the Earth due to neutrino oscillations. Here we have from Table 1 $\alpha = 0,85$ and take $\delta \alpha / \alpha = 0,03.$  

$S_{Th} = S_{T,1} - S_{U,1} - S_{R,1}$ where $S_{R,1}$ is calculated for the place of detector location. The current accuracy of such calculations is 3\%.

Lets estimate the possible accuracy of measurement of values $S_{Th}$, $S_U$ and $\frac{S_{Th}}{S_U}$ where $S_{U} = S_{U,1} + S_{U,2}$ :

\begin{equation}\label{eq:7}
\delta S_{Th} \simeq \sqrt{(\delta S_{T,1})^2 + (\delta S_{U,1})^2} =\sqrt{(\delta S_{T,1})^2 + \alpha^2 (\delta S_{U,2})^2}
\end{equation}

\begin{equation}\label{eq:8}
\delta S_{U,2} = \sqrt{(\delta S_{T,2})^2 + (\delta S_{R,2})^2} \simeq \sqrt{S_{T,2}}
\end{equation}

We can write from (\ref{eq:7}) and (\ref{eq:8}) taking into account that $\alpha$ is near to 1:
\begin{equation}\label{eq:9}
\delta S_{Th} \simeq \sqrt{(\delta S_{T,1})^2 + (\delta S_{T,2})^2} = \sqrt{S_{T,1} + S_{T,2}} = \sqrt{S_T}.
\end{equation} 

We obtained the simple formula to estimate the error:
\begin{equation}\label{eq:10}
\delta \left(\frac{S_{Th}}{S_U}\right) \simeq \frac{\delta S_{Th}}{S_U} \simeq \frac{\sqrt{S_T}}{S_U}.
\end{equation}
\vskip 10mm

\section{Exposition}

We have from (\ref{eq:10}) and (\ref{eq:41}):

 \begin{equation}\label{eq:11}
\delta\frac{S_{Th}}{S_U}(m, t, \eta) \simeq \frac{\sqrt{S_T\cdot m \cdot t \cdot\eta}}{S_{U}\cdot m \cdot t \cdot\eta}
<\frac{1}{3}\cdot\left(\left(\frac{S_{Th}}{S_U}\right)_{BSE} - \left(\frac{S_{Th}}{S_U}\right)_{HE}\right),
\end{equation}
 We can obtain the necessary exposition $m\cdot t$ from resolveing of (\ref{eq:11}) in regard to $m\cdot t$:

\begin{equation}\label{eq:12}
m\cdot t >\frac{9 \cdot S_{T}}{\eta \cdot S_{U}^{2} \cdot \left( \left(\frac{S_{Th}}{S_U}\right)_{BSE} - \left(\frac{S_{Th}}{S_U}\right)_{HE}\right)^2}
\end{equation}

The formula (\ref{eq:12}) gives us the dependance of nesessary exposition on the level of reactors background because  $S_{T} = S_{R} + S_{U} + S_{Th}$.

Lets substitude to (\ref{eq:12}) the values $S_{T}$, $S_{U}$ from Table 3 
for KamLAND laboratory in the case of stopping of all Japanese nuclear power plants and  the values (\ref{eq:4}) and  (\ref{eq:5}) : 

\begin{equation}\label{eq:13}
m\cdot t >\frac{9 \cdot 50}{0,8\cdot 28^2 \cdot (0,28-0,12)^2} = 27.7 kt\cdot y.
\end{equation}

We see from (\ref{eq:13}) that for KamLAND geodetector with feducial mass $m = 5 kt$ is necessary  about 5,5 years of exposition to distinguish between (\ref{eq:4}) and (\ref{eq:5}) in the case of stopping of all Japanese nuclear power plants.

To demonstrate the influence of background from the nuclear power plants  we substitude to (\ref{eq:12}) the values $S_{T}$, $S_{U}$ from Table 4 
for KamLAND laboratory in the case of running of all Japanese nuclear power plants:

\begin{equation}\label{eq:14}
m\cdot t >\frac{9 \cdot 236}{0,8\cdot 28^2 \cdot (0,28-0,12)^2} = 130,7 kt\cdot y.
\end{equation}

Lets take from Table 2 for Sudbury laboratory the values for $S_{T}$, $S_{U}$ and values (\ref{eq:4}) and  (\ref{eq:5}) : 

\begin{equation}\label{eq:15}
m\cdot t >\frac{9 \cdot 99}{0,8\cdot 44^2 \cdot (0,28-0,12)^2} = 22,5 kt\cdot y.
\end{equation}

These calculations shows that it is necessary to build the new generation geoneutrino detector with the fiducial mass not less than 5 kt  and with the background from nuclear power plants not too higher than the signal from U geoneutrino.

\section{Conclusion}

\hspace{4mm} 1. We propose to use the Th/U signal ratio $\frac{S_{Th}}{S_U}$  to distinguish between predictions of Balk Silicat Earth model and Hydridic Earth model.

2. We obtained the simple formula to estimate the error 

of signal ratio (\ref{eq:10}): \hspace{4mm} $\delta \left(\frac{S_{Th}}{S_U}\right) \simeq \frac{\sqrt{S_T}}{S_U}$.

4. We calculated the signals $S_{Th}$ and $S_U$ for Gran-Sasso underground latoratory site, Sudbury Neutrino Observatory site and for Kamioka site (Tables 1, 2, 3).

5. We calculated the signals from the nuclear power plants $S_{R,1}$  and $S_{R,2}$ for Gran-Sasso underground latoratory site, Sudbury Neutrino Observatory site and for Kamioka site (Tables 1, 2, 3).

6. We obtained that for Gran-Sasso underground latoratory site, Sudbury Neutrino Observatory site is necessary the exposition not less than $22 kt \cdot year$ to distinguish between predictions of Bulk Silicate Earth model and Hydridic Earth model (\ref{eq:12}, \ref{eq:13}).

7. We obtained that for Kamioka site is necessary the exposition $27,7 kt \cdot year$ to distinguish between predictions of Bulk Silicate Earth model and Hydridic Earth model (\ref{eq:14}) in the case of stopping of all Japanese nuclear power plants.

8. Our calculations shows that it is necessary to build the new generation geoneutrino detector with the fiducial mass not less than 5 kt  and with the background from nuclear power plants not too higher than the signal from U geoneutrino.

\section{Acknowledgements.}

\hspace{0.5cm}We are grateful to Alexandra Kurlovich, Vladimir Larin, Bayarto Lubsandorzhiev, Stefan Schoenert and Valentina Zavarzina for useful discussions.
This work was supported by grants of Russian Foundation of Basic Research No 13-02-92440 and No 12-02-12124.


\begin{thebibliography}{100}

\bibitem{Borex2013} 
G. Bellini, J. Benziger, D. Bick et al. (Borexino Collaboration) Measurement of geoneutrinos from 1353 days of Borexino. 
arXiv:1303.2571v2 [hep-ex] 4 Apr 2013.

\bibitem{KamLand2013}
A. Gando, Y. Gando, H. Hanakago et al. (The KamLAND Collaboration)  Reactor On-Off Antineutrino Measurement with KamLAND.
arXiv:1303.4667v2 [hep-ex] 20 Mar 2013. 

\bibitem{Mantovani2003} 
Fabio Mantovani, Luigi Garmignani, Gianni Fiorentini, Masello Lissia. Antineutrino from Earth: a reference model and its uncertainties. arXiv:hep-ph/030913v2, 27 November 2003.

\bibitem{lover} J. F. Lovering, J. W. Morgan, Uranium and thorium abundances in stony meteorites: 1. The chondritic meteorites, Journal of Geophysical Research, volume 69, issue 10, 1979,1964.

\bibitem{shinot} Kazunori Shinotsuka, Hiroshi Hidaka and Mitsuru Ebihara, Detailed abundances of rare earth elements, thorium and uranium in chondritic meteorites: An ICP-MS study, Meteoritics, Vol. 30, Issue 6, p.694,1995.

\bibitem{bezruk} Leonid Bezrukov, Geoneutrino and Hydridic Earth mode. Version2. Preprint INR 1378/2014; arXiv:1308.4163[astro-ph.EP].
 
\bibitem{Larin}
Larin,V. N., ed. C. Warren Hunt. Hydridic Earth: the New Geology of Our Primordially Hydrogen-Rich Planet. Polar Publishing, Calgary, Alberta, Canada, 1993, 247p. 

\bibitem{Larin2012}
Herve Toulhoat, Valerie Beaumont, Viacheslav Zgonnik, Nikolay Larin, Vladimir N. Larin. Chemical differentiation of planets: a core issue. 
Aug 2012. 15 pp. e-Print: arXiv:1208.2909 [astro-ph.EP] 

\bibitem{Sinev}
V.V.Sinev. Geoneutrino and the Earth inner parts structure.
Preprint INR 1257/2010. 16 p. June 2010. Moscow. (In Russian).
arXiv:1007.2526v1 [hep-ph] 15 Jul 2010.


\end{thebibliography}
\end{document}